\documentclass{article}

\usepackage[authoryear,round]{natbib} 

\usepackage{graphicx}
\usepackage{amsmath}
\usepackage{amsfonts}
\usepackage{booktabs}
\usepackage{float}

\usepackage{xurl}                  
\usepackage[hidelinks]{hyperref}
\hypersetup{breaklinks=true}


\title{Closing the SNAP Gap: Identifying Under-Enrollment in High-Poverty ZIP Codes}
\author{Auyona Ray}
\date{October 2025}

\begin{document}

\maketitle

\begin{abstract}
   This project began by constructing an index of economic insecurity using multiple socioeconomic indicators. Although poverty alone predicted SNAP participation more accurately than the composite index, its explanatory power was weaker than anticipated, echoing past findings that enrollment cannot be explained by income alone. This led to a shift in focus: identifying ZIP codes with high poverty but unexpectedly low SNAP participation—areas defined here as having a SNAP Gap, where ZIPs fall in the top 30 percent of family poverty and the bottom 10 percent of SNAP enrollment. Using nationally available ZIP-level data from 2014 to 2023, I trained logistic classification models on four interpretable structural indicators: lack of vehicle, lack of internet access, lack of computer access, and percentage of adults with only a high school diploma. The most effective model relies on just two predictors—vehicle access and education—and outperforms tree-based classifiers in both precision and calibration. Results show that economic insecurity is consistently concentrated in rural ZIP codes, with transportation access emerging as the most stable barrier to program take-up. This study provides a nationwide diagnostic framework that can inform the development of scalable screening tools for targeting outreach and improving benefit access in underserved communities.
\end{abstract}

\clearpage

\section{Introduction}
The Supplemental Nutrition Assistance Program (SNAP) is the largest federal nutrition assistance program in the United States \cite{cbpp2022}, yet millions of eligible households remain unenrolled each year \cite{tiehen2013,suttles2024}. Understanding where and why these gaps occur is critical for improving program delivery. While prior research has emphasized the relationship between poverty and SNAP participation \cite{almond2011,hoynes2016}, the persistence of under-enrollment in certain communities suggests the presence of structural barriers beyond income alone \cite{berkowitz2017,ajph2019}.

To address this problem, I introduce the concept of the \textit{SNAP Gap}: ZIP codes that are simultaneously in the top 30 percent of family poverty and the bottom 10 percent of SNAP participation. These areas represent communities where high need does not translate into proportional program take-up. For clarity, I refer to such ZIP codes as \textit{fragile}, and use the term \textit{prevalence} to describe the share of ZIPs that meet this definition in a given period.

This paper advances the study of program under-enrollment by demonstrating that a simple, interpretable logistic framework can consistently diagnose these anomalies nationwide. Specifically, I show that two structural indicators—vehicle access \cite{saia2023} and educational attainment \cite{hoynes2016}—explain SNAP under-enrollment more effectively than composite indices or digital access measures \cite{strover2017,graves2021,pew2021}. The findings suggest that structural barriers, particularly limited transportation and lower educational attainment, are more predictive of under-enrollment than factors such as internet or computer access. This reframes the challenge of benefit delivery as one rooted in mobility and human capital, rather than eligibility or awareness alone \cite{suttles2024}.

\section*{Research Foundation: Theoretical and Empirical Background on the SNAP Participation Gap}

The Supplemental Nutrition Assistance Program (SNAP) delivers benefits that extend far beyond its core function of reducing food insecurity. A growing body of research demonstrates that early-life access to SNAP leads to sustained improvements in health, educational attainment, and long-term economic stability---particularly among vulnerable populations \cite{hoynes2016,almond2011}. These individual-level gains scale up to broader macroeconomic and societal returns, positioning SNAP not merely as a safety-net program but as a foundational investment in human capital and public infrastructure \cite{tiehen2013,cbpp2022}.

Early exposure to SNAP has been linked to improved adult well-being across multiple dimensions. Hoynes, Schanzenbach, and Almond (2016) show that individuals who grew up in counties with greater SNAP availability in the 1960s and 1970s experienced lower rates of metabolic syndrome, higher educational attainment, and increased adult earnings---especially among women \cite{hoynes2016}. Complementing this, Almond, Hoynes, and Schanzenbach (2011) find that prenatal access to food stamps increased infant birth weight, particularly among low-income mothers, providing a biological pathway through which early-life nutritional support enhances long-term health \cite{almond2011}.

These individual benefits generate measurable returns at the societal level. Tiehen, Jolliffe, and Smeeding (2013) estimate that SNAP reduces overall poverty by 14--16 percent and cuts deep poverty by more than 50 percent among non-elderly households \cite{tiehen2013}. Berkowitz et al. (2017) expand this lens to the healthcare system, reporting that adult SNAP participants incur approximately \$1,400 less in annual medical expenses compared to income-eligible nonparticipants \cite{berkowitz2017}. This finding reframes SNAP as a form of preventive public-health infrastructure, with downstream fiscal dividends rather than merely transfer-based spending.

Population-level studies reinforce these conclusions. Adults aged 40--64 who receive SNAP have 1--2 percent lower all-cause mortality than comparable nonparticipants \cite{ajph2019}. Meanwhile, severe food insecurity is associated with a 6.2-year reduction in life expectancy at age 50 \cite{jama2023}. Together, these findings demonstrate that SNAP not only extends lives but also reduces long-run public health expenditures, positioning enrollment as a driver of national productivity and longevity.

SNAP’s spillover effects extend to public safety and community stability. Monthly benefit cycles that smooth household consumption reduce theft by roughly 20 percent \cite{rest2019}, and reinstating eligibility for formerly incarcerated individuals has been shown to lower rates of recidivism \cite{psj2018}. These outcomes underscore SNAP’s public-goods function, linking food access to safer, more economically stable communities.

The fiscal multipliers further reinforce the program’s long-horizon value. Analyses by the Center on Budget and Policy Priorities (2022) estimate that every \$1 in SNAP benefits generates \$1.50--\$2.00 in economic activity, and as much as \$62 in lifetime social returns for children \cite{cbpp2022}. From a human-capital perspective, these multipliers reflect the compounding returns of early-life nutritional adequacy---improving cognitive development, educational persistence, and long-term earnings trajectories. Conversely, program withdrawal or non-participation erodes these outcomes.

Persistent under-enrollment---commonly referred to as the ``SNAP Gap''---thus represents not merely an administrative inefficiency but a structural loss of future human capital. Communities with high rates of unenrolled eligible households face compounding deficits: higher healthcare costs, lower educational outcomes, reduced labor productivity, and shorter life expectancy. Understanding the roots of this erosion requires a shift from individualized behavioral explanations toward structural determinants of access---specifically, the roles of mobility, education, and digital connectivity in shaping program participation \cite{suttles2024,saia2023,strover2017}.

Although the literature consistently demonstrates SNAP’s positive impacts, some critiques highlight limitations that can be reframed as additional structural barriers to access. For example, inconsistent state-level implementation, administrative discretion, and variability in outreach efforts have been shown to shape local enrollment outcomes \cite{suttles2024,ssm2023}. Likewise, reduced effects among mixed-status households or transient populations may reflect not a failure of SNAP per se, but the exclusionary conditions under which it operates \cite{ssm2023}. Methodological debates---such as observational versus quasi-experimental designs---further point to how structural and contextual differences complicate national generalizations. Rather than contradicting the program’s value, these findings emphasize the need to understand under-enrollment as a product of fragmented delivery systems, uneven institutional trust, and localized structural disadvantage.

The following subsections examine these structural constraints in greater depth.

\subsection*{Structural and Administrative Barriers to Enrollment}

Despite SNAP’s well-documented long-term benefits, a substantial share of eligible households remain unenrolled---an enduring gap that points to structural and administrative frictions in program access. Much of the literature attributes under-enrollment to education and bureaucratic literacy: an individual's ability to navigate complex systems of eligibility verification, documentation, and digital submission \cite{hoynes2016,suttles2024}. Here, educational attainment functions not simply as a measure of human capital but as a proxy for procedural fluency---the capacity to understand program criteria, complete online forms, and comply with time-sensitive requirements.

Recent state-level data from Indiana illustrates this link between education and enrollment outcomes. Suttles, Babb, and Knudsen (2024) analyze administrative records and find that communities with lower average education levels face significantly longer processing times, higher denial rates, and greater difficulty completing required interviews \cite{suttles2024}. Their quantile regression results suggest that these inefficiencies stem less from individual applicant behavior and more from broader educational and socioeconomic contexts. The implication is clear: education acts as a structural determinant of administrative capacity, shaping how households interact with and navigate program bureaucracy.

Administrative complexity compounds other barriers, including documentation burdens, policy inconsistency, and stigma. Research in Social Science \& Medicine (2023) finds that SNAP participation among Latino households is influenced by local immigration enforcement climates---particularly among eligible noncitizens and mixed-status families \cite{ssm2023}. These findings expand the definition of ``administrative burden'' beyond procedural friction to include the psychosocial costs of engaging with state systems, especially in environments where institutional trust is low.

Education also intersects with demographic and geographic factors that obscure accurate assessments of program participation. A 2024 Journal of the Academy of Nutrition and Dietetics (JAND) evaluation of SNAP-Ed programs for college students highlights how definitional mismatches---such as household-based eligibility metrics applied to student populations---can distort local enrollment data \cite{jand2024}. College-town ZIP codes, for instance, may appear to have unusually low take-up rates, not due to program failure, but because student eligibility criteria diverge from those applied to families. Such examples underscore the importance of contextualizing participation statistics within demographic and educational landscapes.

Collectively, these strands of evidence position education as both a capacity and context variable. It shapes not only individual-level engagement with the application process but also the broader informational and institutional environment that supports---or impedes---enrollment. Communities with lower rates of high school completion or limited access to postsecondary education often lack the bureaucratic literacy and civic infrastructure necessary to translate eligibility into participation. These dynamics justify the inclusion of education in this study as a structural variable: a community-level marker of human capital infrastructure, essential for bridging the gap between policy access and policy uptake.

\subsection*{Mobility Constraints: Transportation and Vehicle Access}

Beyond educational and administrative hurdles, mobility limitations represent a persistent and underappreciated structural barrier to SNAP participation. For many eligible households, the lack of a reliable vehicle---or limited access to affordable public transportation---creates logistical constraints that impede both enrollment and benefit utilization. Physical distance from welfare offices, authorized grocery retailers, and community organizations imposes higher opportunity costs on low-income individuals, particularly those managing unstable work schedules, childcare responsibilities, or restricted time flexibility \cite{saia2023,harnack2019}.

Saia (2023) identifies transportation insecurity as one of the most consistent predictors of food insecurity, even after adjusting for income \cite{saia2023}. Rural residents and low-income suburban households are especially disadvantaged, as longer travel distances and sparse public transit networks compound the time and financial burden of program engagement \cite{harnack2019}. These findings underscore that transportation is not merely a logistical convenience but a structural prerequisite for effective participation in programs that require in-person verification or regular food access.

Policy responses have begun to mitigate these spatial constraints through technological innovations, most notably the introduction of online grocery purchasing using Electronic Benefit Transfer (EBT) cards. Martinez et al. (2018) provide early mixed-methods evidence suggesting this approach could partially address the ``no-vehicle problem'' \cite{martinez2018}. However, uptake among SNAP recipients remains limited by digital literacy gaps, delivery fees, and uneven participation by online retailers. Similarly, findings from the 2023 Double Up Food Bucks clinic evaluation highlight that while participants appreciated the potential for healthier food options, lack of transportation and limited availability of nearby markets remained dominant deterrents \cite{phn2023}.

These studies point to a layered problem: even where technological alternatives exist, their impact is moderated by underlying structural inequalities---namely digital access, retail proximity, and transportation infrastructure \cite{strover2017,graves2021,pew2021,pew2022}. In this context, vehicle ownership emerges as a foundational variable, shaping not only enrollment likelihood but also the practical value of participation \cite{saia2023,harnack2019}. Households without vehicles face compounded barriers: reduced access to administrative offices, fewer retail options, and limited flexibility to redeem benefits.

\subsection*{The Digital Divide as a Time-Variant Barrier}

Early research on SNAP enrollment in the 2010s frequently treated the digital divide as a fixed structural limitation. Households without reliable broadband, mobile devices, or basic digital skills lacked access to essential information about the program and were often unable to complete online applications or interact with digital systems. Strover et al. (2017) and Horrigan (2016) documented wide disparities in internet access and digital literacy, particularly among rural and low-income populations \cite{strover2017,horrigan2016}.

However, more recent evidence suggests that these digital access gaps have narrowed considerably. Graves et al. (2021) report rising rates of device ownership and improved connectivity among younger, lower-income populations \cite{graves2021}, while Pew Research (2021) finds that broadband adoption surged across most demographic groups during the COVID-19 pandemic \cite{pew2021}. Federal and state infrastructure investments further accelerated this trend: EverythingPolicy (2023) documents a substantial reduction in the rural-urban connectivity gap after 2019, although disparities in speed, reliability, and affordability remain \cite{everythingpolicy2023}.

This evolving landscape repositions the digital divide as a time-variant barrier---one that was sharply constraining in earlier years but has become less predictive of access inequality in more recent periods. For the current study, this shift has two core implications. First, digital access no longer accounts for non-participation as strongly as it did a decade ago. Second, this decline highlights which barriers have not improved---namely, transportation and education---which now exert greater relative influence on SNAP enrollment outcomes.

\subsection*{Rural Amplification of Structural Barriers}

While education, mobility, and digital connectivity each constrain SNAP participation in distinct ways, their effects are often amplified in rural settings, where geographic isolation intersects with limited institutional infrastructure. Rural areas typically exhibit higher SNAP eligibility rates but lower participation relative to urban centers---a paradox rooted in overlapping deficits in transportation, human capital, and service access. This concentration of disadvantage makes the SNAP Gap not merely a socioeconomic issue, but also a spatial one.

Harnack et al. (2019) emphasize that rural households encounter structural food-access barriers that urban-focused models frequently overlook \cite{harnack2019}. Longer travel distances to administrative offices and authorized retailers---combined with higher fuel costs and limited public transit---intensify the impact of vehicle scarcity. Saia (2023) reinforces this finding, showing that mobility constraints disproportionately burden rural households, highlighting transportation as both a logistical and economic barrier to participation \cite{saia2023}.

Educational disparities further compound these challenges. Rural communities generally have lower rates of postsecondary attainment and a higher proportion of residents with only a high school diploma or less \cite{hoynes2016}. This human-capital profile limits bureaucratic literacy---the ability to complete digital or paper applications, navigate documentation requirements, and resolve benefit interruptions.

Although recent years have seen measurable progress in narrowing the digital divide, that progress has not been evenly distributed. Pew Research Center (2022) reports that rural broadband access still trails urban and suburban levels, with affordability cited as a primary barrier \cite{pew2022}. Even where broadband is technically available, slower speeds and higher service costs constrain the usability of online SNAP platforms. As a result, rural residents remain disproportionately dependent on in-person processes---precisely those most affected by transportation and education constraints.

\subsection*{Synthesis and Research Contribution}

The preceding literature demonstrates that the Supplemental Nutrition Assistance Program (SNAP) delivers significant long-term economic, health, and social returns, yet continues to face persistent under-enrollment in many eligible communities \cite{tiehen2013,berkowitz2017,ajph2019}. Taken together, the evidence suggests that participation is shaped less by individual choice or short-term policy dynamics than by structural constraints---conditions embedded in the social, spatial, and institutional contexts of local areas.

This synthesis yields two core insights that guide the present study. First, the most enduring forms of the SNAP Gap stem from material and educational infrastructures rather than from informational deficits. Vehicle access and educational attainment reflect the tangible and cognitive resources that households must possess to convert eligibility into participation---resources that cannot be substituted by digital platforms alone \cite{saia2023,hoynes2016}. Second, as digital connectivity has improved while transportation and educational disparities have remained relatively stable, the explanatory power of the latter should have increased over time \cite{graves2021,pew2021,pew2022}.

Despite the growing recognition of these structural constraints, there remains a clear empirical gap: few studies have systematically identified where SNAP under-enrollment is most pronounced using scalable, interpretable indicators at fine-grained geographic levels \cite{suttles2024}. Much of the existing work focuses on individual or household characteristics, or state-level policy variations, without providing a ZIP-level diagnostic tool for isolating fragile communities---those with high poverty but unexpectedly low take-up.

Accordingly, this study makes three contributions to the existing literature:
\begin{enumerate}
    \item It provides a comprehensive, ZIP-level analysis of the structural determinants of SNAP under-enrollment, integrating education, vehicle availability, and digital access within a unified empirical framework.
    \item It offers a longitudinal perspective by comparing two time periods---before and after the expansion of digital infrastructure---thereby assessing how the salience of these structural barriers has changed over time.
    \item It highlights the continued rural persistence of these constraints, advancing understanding of how geography mediates access to social policy.
\end{enumerate}

Together, these contributions reframe the SNAP Gap not as a behavioral anomaly, but as the outcome of uneven infrastructure and human capital investment---setting the stage for the empirical analysis that follows.

\subsection*{Motivation}
\label{sec:motivation}
This project began with the goal of developing a ZIP-level “fragility index” to identify economically vulnerable communities across the United States. I define fragility as a condition in which a ZIP code shows structural indicators of economic insecurity beyond poverty alone—for example, high unemployment, low income, or cost-burdened renters \cite{tiehen2013}. Using American Community Survey data (2014–2018), I combined these variables through Principal Component Analysis to create a composite fragility score.

Validation against Supplemental Nutrition Assistance Program (SNAP) participation, however, showed that poverty alone was a stronger predictor than the composite index \cite{almond2011,hoynes2016}. This raised questions about the usefulness of fragility as a broad construct. More importantly, it revealed a distinct set of cases: ZIP codes with high poverty but unexpectedly low SNAP enrollment. I define these communities as exhibiting a SNAP Gap, where ZIPs fall in the top 30 percent of family poverty and the bottom 10 percent of SNAP participation \cite{suttles2024}.

Reframing the project around the SNAP Gap shifted the emphasis from measuring general economic strain to diagnosing under-enrollment directly. This pivot clarified the study’s contribution: rather than diluting the poverty signal with weaker variables, the analysis isolates communities that appear to face structural barriers to accessing benefits despite high economic need.

\subsection*{Organization of the Paper}
The remainder of this paper is organized as follows. Section~2 describes the data sources and construction of variables. Section~3 outlines the conceptual framework and defines key terms, including fragility, prevalence, and the SNAP Gap. Section~3 also presents the modeling approach and evaluation metrics. Section~4 reports the results, highlighting key predictors of SNAP under-enrollment. Finally, Section~5 concludes with policy implications and directions for future research.

\section{Data}
\label{sec:data}

\subsection{Primary Sources}
The analysis uses ZIP Code Tabulation Area (ZCTA)–level data products from
PolicyMap, HUD, and supplemental reference files \cite{everythingpolicy2023}. Data include family poverty counts, 
family SNAP counts, contextual ``Plus 4'' socioeconomic indicators, and a HUD/USPS ZIP–tract crosswalk 
to derive rural/urban designations \cite{strover2017,harnack2019}.

\subsection{Variable Construction}
Key variables are defined as follows:
\begin{itemize}
    \item \textbf{Family poverty rate} and \textbf{family SNAP count}.
    \item Uptake ratio: 
    \[
    s_z = \frac{\text{SNAP families}_z}{\text{poverty families}_z}, \quad s_z \leq 1.0.
    \]
    \item Poverty floor: $p_z \geq 0.15$.
    \item Quantile thresholds: high-poverty $\tau_{\text{hi}} = \text{70th percentile of } p_z$,
    low-uptake $\tau_{\text{lo}} = \text{10th percentile of } s_z$.
    \item Target variable $y_z$: flag set to 1 if 
    $p_z \geq \tau_{\text{hi}}$ and $s_z \leq \tau_{\text{lo}}$.
    \noindent
    \textit{Note:} PolicyMap provides both family-level and total poverty counts, but only 
family-level SNAP counts are available. To keep numerator and denominator aligned with the same 
population unit (households/families), all analyses use family-level variables.\footnote{This also avoids 
known mismatches in college-town or group-quarters contexts where individual rates diverge from family units \cite{jand2024}.}
\end{itemize}

\noindent
\textit{Note:} PolicyMap provides both family-level and total poverty counts, but only 
family-level SNAP counts are available. For consistency, all analyses use family-level variables.

\subsection{Limitations}
Additional details on data limitations, including PolicyMap compositing, area-type assumptions, and missingness effects, are provided in Appendix~\ref{sec:appendix-area} \cite{everythingpolicy2023,strover2017,harnack2019}.

\noindent
After eligibility filters and cleaning, the analytic sample comprises 
6,802 ZIPs in P1 and 5,466 ZIPs in P2.

\section{Methods}
\label{sec:methods}

\subsection{Design Rationale and Modeling Choices}
\label{sec:design-rationale}

Our design choices are guided by two goals: (i) produce a screening tool that an agency can 
operate prospectively, and (ii) keep the model compact, interpretable, and portable across time 
and geographies.

\paragraph{Eligibility floor and quantiles.}
We impose a family-poverty floor of $\underline{p}=0.15$ to focus attention on ZIP codes where 
poverty exposure is unambiguously material. The 15\% cutoff follows the policy dialogue on 
``persistent poverty'' \cite{tiehen2013} and keeps eligibility broad enough to be operationally relevant while 
excluding noise in very low-poverty areas. Within the eligible pool $\mathcal{E}$, the target is 
defined with \emph{quantile} thresholds---top 30\% in poverty and bottom 10\% in uptake. 
Quantiles are scale-free: they adapt to period- and area-specific distributions without baking in 
arbitrary absolute levels, and they preserve comparability across P1$\rightarrow$P2 and across 
urban/rural mixes. In short, the rule flags ``high-need, low-uptake'' ZIPs no matter where the 
overall level sits in a given window \cite{suttles2024}.

\paragraph{Why a binary target (not continuous regression)?}
The operational question for outreach is binary---\emph{does this ZIP warrant attention now?} 
A classification target yields ranked lists, prevalence-anchored decisions, and precision@K 
metrics that map directly to budgeted field capacity. We still analyze OLS residuals of 
SNAP-on-poverty as a diagnostic complement to catch idiosyncratic under-uptake that the 
quantile rule may miss, but deployment needs a list rather than a predicted ratio \cite{almond2011,hoynes2016}.

\paragraph{Predictor set and tautology avoidance.}
Poverty and SNAP \emph{define} the target, so we exclude them as features to avoid learning back 
our own rule. Instead we use four ``Plus4'' structural indicators---no vehicle, no internet, no 
computer, and high-school-only attainment---that capture transportation, digital, and human-capital 
barriers plausibly upstream of program take-up \cite{saia2023,strover2017,graves2021,hoynes2016}. 
This keeps the model interpretable and portable: agencies can score future data with inputs they 
actually maintain. A richer set (race, unemployment, rent burden) is a natural extension, but we 
prioritize a uniform, reproducible baseline across P1 and P2.

\paragraph{Anomalies, unknowns, and fixed area types.}
ZIPs with SNAP $>$ poverty are \emph{retained} and flagged. Removing them would risk discarding 
real high-uptake outliers alongside data quirks, and could make performance look artificially strong. 
Similarly, ZIPs whose tract mixes are ``Unknown'' are shown descriptively but excluded from model 
fitting, because many are non-residential or PO-box geographies that would dilute calibrated thresholds. 
Area designations (Urban/Mixed/Rural) are fixed from the 2019 HUD crosswalk \cite{everythingpolicy2023}; 
annual reweighting at ZIP level is not available, and we assume these mixes do not swing sharply between P1 and P2.

\paragraph{Model family, calibration, and decision threshold.}
We benchmark a transparent logistic model against random forest and gradient boosting to test for 
useful nonlinearity. With few, correlated predictors and a very rare positive class, simple models 
perform best and remain explainable. Probabilities are calibrated with isotonic regression---a 
monotone, shape-flexible correction well-suited to under/over-confident scores---rather than Platt 
scaling, which assumes a sigmoid. For decisions, we anchor the cutpoint to training prevalence 
$\pi_{\text{P1}}$, which keeps flagged counts commensurate with historical incidence and avoids the 
``optimal'' ROC/F1 thresholds that can explode selections under heavy class imbalance.

\paragraph{Stratifying by settlement type.}
Program access barriers plausibly differ with settlement type (distance and transit in rural places; 
digital deserts and administrative frictions in urban ones). We therefore recompute quantiles within 
Urban/Mixed/Rural subsets and re-estimate models in each. This reveals where each structural barrier 
carries signal and prevents one context from dominating thresholds for the other \cite{harnack2019,saia2023}.

\paragraph{Evaluation emphasis.}
Average Precision and precision@K take precedence over ROC AUC because the planning problem is 
``how many true gaps do we catch in the top 1--5\% of the list we can actually visit?'' 
We evaluate strictly out-of-time on P2 to mimic deployment and to expose temporal drift that 
cross-validation within P1 would miss \cite{suttles2024}. Residual diagnostics accompany the quantile rule to 
identify edge cases and alternative narratives of under-enrollment.

\paragraph{What the results imply.}
Transportation (no vehicle) and education (HS-diploma-only) carry the clearest, most stable signal 
in out-of-time tests, with digital access gaining strength in urban and mixed areas but looking 
muted in pooled samples \cite{saia2023,hoynes2016,strover2017,graves2021}. That pattern likely reflects heterogeneity: 
digital barriers matter where eligibility is high and density amplifies administrative frictions; 
in sparse rural areas, physical access dominates. Together, these reveal actionable levers for 
targeted outreach and future data collection.

\subsection{Analytic Framework}
The pipeline proceeds in four stages: (1) cleaning and eligibility filtering, 
(2) target construction, (3) model training on P1 (2014--2018), 
and (4) out-of-time evaluation on P2 (2019--2023). 
This forward design emulates prospective policy use: all models are trained on 
historical data and evaluated strictly on later periods, ensuring that no 
information from the test window leaks into training. Such temporal separation is 
essential when the motivating question is whether fragile ZIP codes could be 
identified in real time to guide outreach \cite{suttles2024}.

\subsection{Hypotheses}
The analysis begins with a structural definition of fragility. 
Let $p_z \in [0,1]$ be the family poverty rate and
$s_z = \text{SNAP}_z / \text{poverty}_z \in [0,1]$ the SNAP uptake ratio.
To restrict to contexts with meaningful poverty exposure, a floor of 
$\underline{p}=0.15$ is imposed. The eligible set is therefore
\[
\mathcal{E} = \{ z : p_z \ge \underline{p},\; p_z > 0,\; s_z \text{ finite} \}.
\]

Within this set, quantile cutpoints are defined by
\[
\tau_{\text{hi}} = \operatorname{quantile}_{0.70}\{p_z : z \in \mathcal{E}\},\qquad
\tau_{\text{lo}} = \operatorname{quantile}_{0.10}\{s_z : z \in \mathcal{E}\}.
\]
The thresholds select ZIP codes simultaneously in the top 30\% of poverty and 
the bottom 10\% of uptake \cite{tiehen2013}.

\textbf{Hypothesis (fragility):} a ZIP is fragile when
\[
p_z \ge \tau_{\text{hi}} \quad \text{and} \quad s_z \le \tau_{\text{lo}}.
\]

\paragraph{Residual diagnostics.}
To assess whether the quantile-based rule might miss unusual cases, 
ordinary least squares regression of SNAP counts on poverty counts is estimated:
\[
\hat{\alpha}, \hat{\beta} = \arg\min_{\alpha,\beta} 
\sum_{z \in \mathcal{E}} (\text{SNAP}_z - \alpha - \beta\,p_z )^2.
\]
Residuals $\hat{\epsilon}_z = \text{SNAP}_z - \hat{\alpha} - \hat{\beta}p_z$
quantify departures from expected uptake. Large negative values suggest under-enrollment 
not explained by the quantile thresholds. This provides a diagnostic complement 
rather than a competing definition \cite{almond2011,hoynes2016}.

\subsection{Target Variable}
The binary target $y_z$ is defined as
\[
y_z =
\begin{cases}
1, & \text{if } p_z \ge \tau_{\text{hi}} \ \text{and}\ s_z \le \tau_{\text{lo}}, \\
0, & \text{otherwise (within } \mathcal{E}), \\
\text{NA}, & \text{if } z \notin \mathcal{E}.
\end{cases}
\]
This construction balances tractability with policy interpretability: a value of 1 
means a ZIP has both high poverty and anomalously low uptake. NA values ensure 
that ZIPs failing eligibility filters are explicitly excluded rather than 
misclassified \cite{suttles2024}.

\subsection{Predictor Variables}
Predictors are the four ``Plus4'' socioeconomic indicators from PolicyMap:
\[
x_z = \bigl(
\text{\% no vehicle},\;
\text{\% no internet},\;
\text{\% no computer},\;
\text{\% HS diploma only}
\bigr).
\]
These variables represent structural barriers to program access: transportation, 
digital connectivity, and educational attainment \cite{saia2023,strover2017,graves2021,hoynes2016}. 
Poverty and uptake do not enter as predictors, but only through the target definition, to avoid tautological models. 
Multivariate experiments consider all nonempty subsets of these four predictors.

\subsection{Model Training and Evaluation Metrics}
The study benchmarks three interpretable classifiers (logistic regression, random forest, and gradient boosting) using stratified cross-validation on the training period (P1) and out-of-time testing on P2. Logistic regression, calibrated by isotonic regression, yields probability scores that are converted into binary classifications using a prevalence-anchored cutoff based on historical SNAP Gap incidence \cite{suttles2024}. Performance is assessed through precision–recall metrics, average precision, and calibration diagnostics, emphasizing top-rank accuracy rather than overall accuracy due to class imbalance.

Models are estimated separately for Urban, Rural, and Mixed ZIPs to account for contextual variation in fragility, with quantile thresholds and prevalence levels recalculated for each subset \cite{harnack2019,saia2023}. Temporal partitioning (2014–2023) ensures comparability across periods. Fixed thresholds (70th-percentile poverty, 10th-percentile uptake) and a 15 \% poverty floor maintain consistent definitions across years \cite{tiehen2013}.

\subsection{Area Context and Stratified Modeling}
Because SNAP participation barriers differ by settlement type, 
all models were stratified by Urban, Rural, and Mixed ZIP codes. 
Thresholds for high poverty and low uptake were recalculated within each subset 
to ensure local comparability. Fragility was consistently concentrated in 
Rural ZIPs, which accounted for roughly three quarters of low-uptake areas \cite{harnack2019}. 
Full area-specific thresholds, prevalence values, and permutation importances 
are provided in Appendix~\ref{sec:appendix-area}.

\section{Results}
\label{sec:results}

\noindent
\textbf{Reading the results.} We first summarize descriptive patterns (heterogeneity of uptake 
conditional on poverty and the geography of low-uptake tails), then report out-of-time model 
performance with an emphasis on precision in the top of the ranked list. We close with calibration 
and residual diagnostics that distinguish systemic from idiosyncratic fragility.

\subsection{Descriptive and Exploratory Analysis}
Initial exploration confirmed substantial heterogeneity in uptake conditional on poverty. 
Scatterplots of SNAP vs.\ poverty showed strong linear association overall, 
but with wide dispersion—suggesting that some ZIPs fall systematically 
below expected levels of enrollment \cite{almond2011,hoynes2016}. 
Histograms of OLS residuals reinforced 
this view: the distribution was heavily left-skewed, implying a subset of 
ZIPs with unexplained under-uptake.

\begin{figure}[ht]
    \centering
    \includegraphics[width=0.65\linewidth]{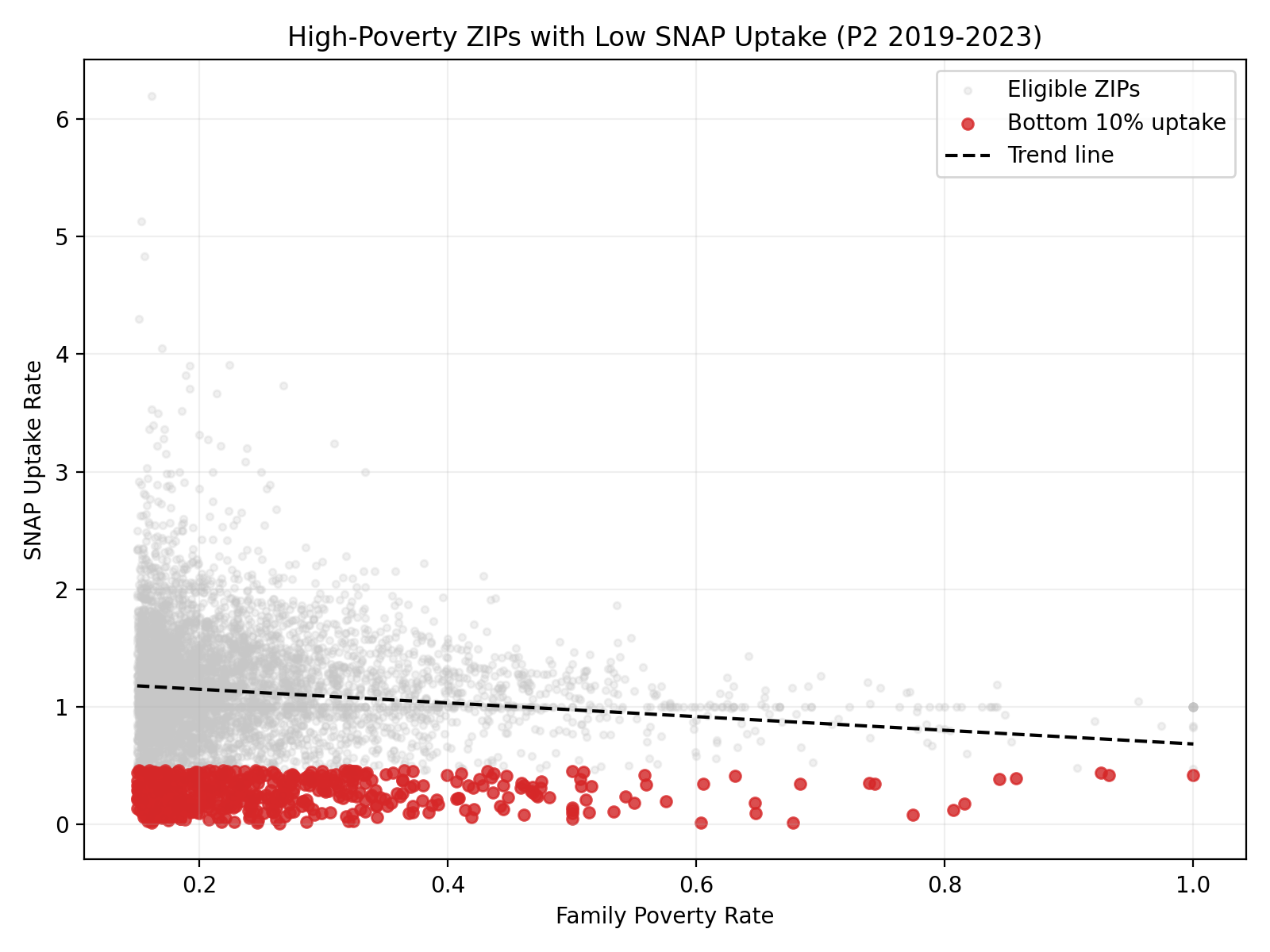}
    \caption{Scatterplot of family SNAP counts versus family poverty counts.
A strong linear trend coexists with wide vertical dispersion: many ZIPs sit well below the 
fitted line given their poverty counts, consistent with pockets of under-enrollment that 
predictive models and residual checks aim to identify.}
    \label{fig:scatter_snap_poverty}
\end{figure}

\begin{figure}[ht]
    \centering
    \includegraphics[width=0.55\linewidth]{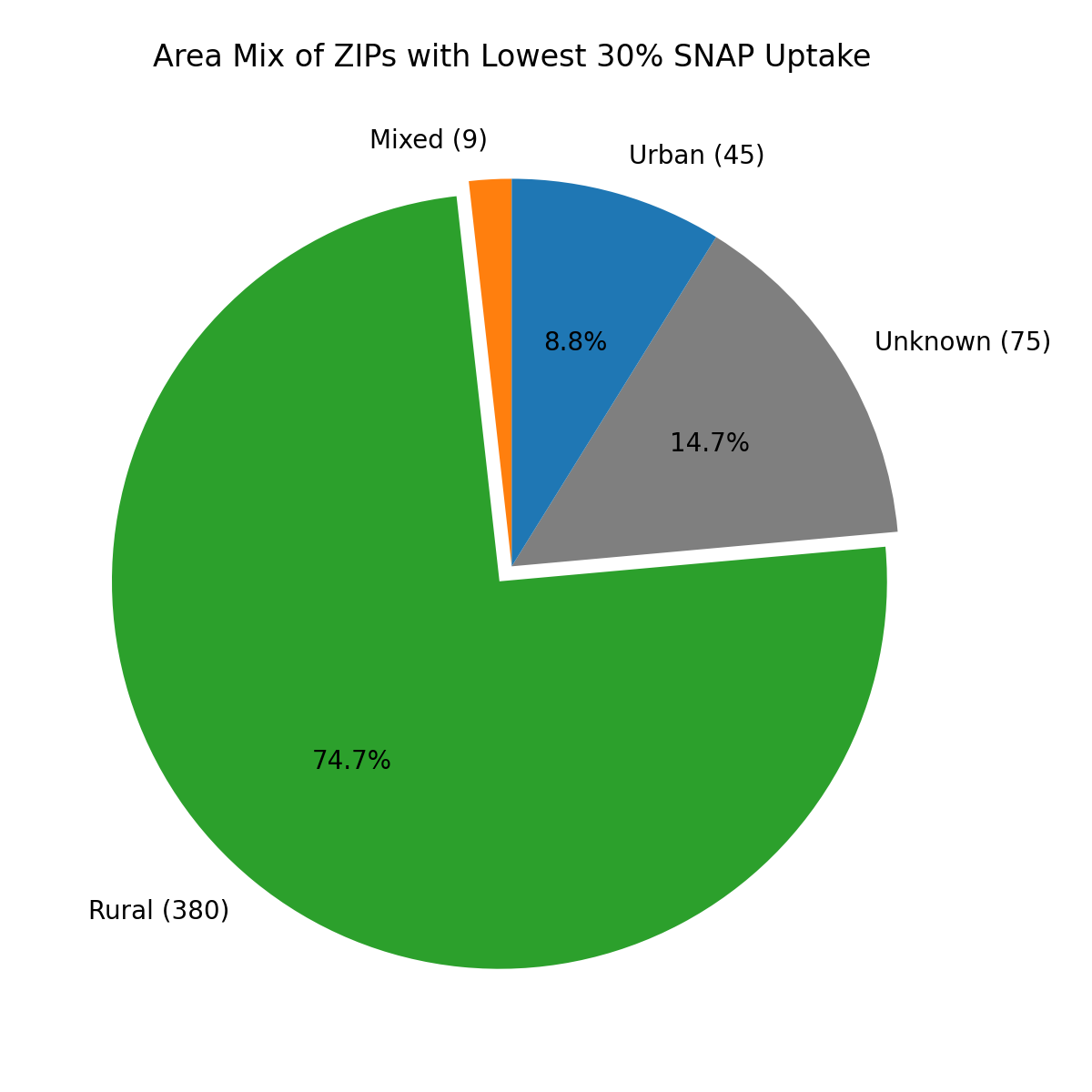}
    \caption{Distribution of area types among Bottom 30\% low-uptake ZIPs.
Rural ZIPs dominate the fragile tail even after adjusting thresholds within areas, reinforcing 
the need for rural-specific thresholds and interventions.}
    \label{fig:pie_bottom30}
\end{figure}

Quantile classifications highlighted the geography of fragility. Bottom 10\% 
and 30\% uptake thresholds identified ZIPs disproportionately concentrated in 
rural settings, even after adjusting for poverty levels \cite{harnack2019,saia2023}. 
Disaggregation by urban, mixed, and rural categories confirmed that fragile areas are not 
uniformly distributed. Figure~\ref{fig:pie_bottom30} provides a visual 
summary of this distribution.

\subsection{Univariate Logistic Models}
Univariate logistic regressions, trained on P1 and evaluated on P2, 
produced the results in Table~\ref{tab:univariate}. 
Because fragility is rare ($\pi_{\text{P1}} \approx 3.1\%$, 
$\pi_{\text{P2}} \approx 3.8\%$), simple predictors often performed poorly. 
The vehicle variable stood out: it achieved AUC near 70 and AP near 8.5, 
far outperforming the other three features \cite{saia2023}.

\begin{table}[ht]
\centering
\caption{Univariate logistic regression performance on P2. Dashes (--) indicate values near zero. All values are percentages.}
\label{tab:univariate}
\resizebox{\textwidth}{!}{%
\begin{tabular}{lccccccc}
\hline
Predictor & AUC & AP & Precision & Recall & $F_1$ & Accuracy & Pre@1\% / Pre@5\% \\
\hline
\% no vehicle      & 69.5 & 8.5 & 11.8 & 36.5 & 17.8 & 87.2 & 14.8 / 12.8 \\
\% HS diploma only & 59.8 & 6.4 & 11.8 & 15.9 & 13.6 & 92.3 & -- / -- \\
\% no computer     & 57.5 & 5.3 & 2.0  & 0.5  & 0.8  & 95.3 & -- / -- \\
\% no internet     & 53.2 & 4.5 & 5.6  & 1.4  & 2.3  & 95.3 & -- / -- \\
\hline
\end{tabular}%
}
\end{table}

The recall for vehicle access (36.5\%) was especially notable: 
while absolute precision was modest, this predictor alone identified 
over a third of fragile ZIPs. Other predictors, particularly computer and 
internet access, provided little discrimination when used individually \cite{strover2017,graves2021}.

\subsection{Multivariate Logistic Models}
Multivariate combinations modestly improved performance. The best specification, 
combining vehicle access and education, yielded AUC $\approx 71.1$ and AP $\approx 9.8$, 
outperforming all univariate models (Table~\ref{tab:multivariate}). 
Coefficient estimates indicated a negative coefficient for vehicle access 
($\hat{\beta}=-0.3125$), supporting the substantive interpretation that 
household vehicle ownership is protective against fragility \cite{saia2023,hoynes2016}. 

\begin{table}[ht]
\centering
\resizebox{\textwidth}{!}{%
\begin{tabular}{lccccccc}
\hline
Predictors & AUC & AP & Precision & Recall & $F_1$ & Accuracy & Pre@1\% / Pre@5\% \\
\hline
\% no vehicle + \% HS diploma only & 71.1 & 9.8 & 14.1 & 20.2 & 16.6 & 92.3 & 14.8 / 13.6 \\
All four predictors                & 69.1 & 8.6 & 11.8 & 36.5 & 17.8 & 87.2 & 9.3 / 12.8 \\
\hline
\end{tabular}}
\caption{Multivariate logistic regression results (P2).}
\label{tab:multivariate}
\end{table}

Interestingly, the full four-variable model offered little gain relative to 
vehicle alone. This suggests that fragility is most strongly captured by 
transportation and education, while digital variables add limited incremental 
signal at ZIP scale \cite{strover2017,graves2021,pew2021}.

\subsection{Tree-Based Benchmarks}
Tree-based methods were tested as nonlinear alternatives. Random forests with 
all predictors achieved AUC of 66.3 and AP of 6.4, underperforming logistic regression. 
Gradient boosting raised AUC modestly to 68.0 and AP to 7.5, but still fell short 
of the best logistic results.\footnote{Exact values varied by hyperparameter 
configuration; the strongest gradient boosting models clustered around AUC 68, AP 7.5.}
Permutation importances highlighted internet access and education as top predictors, 
aligning with logistic regression’s interpretation of educational barriers \cite{suttles2024}. 
Overall, tree-based methods did not offer meaningful advantages given the small 
positive class and limited predictor set.

\subsection{Area-Specific Results}
Area-stratified analyses reinforce the role of geography. After re-computing
area-specific thresholds $(\tau_{\text{hi}}, \tau_{\text{lo}})$, the All-sample
P2 holdout ($n \approx 25{,}241$) shows internet access (\% no internet) and
education (\% HS diploma only) as the strongest pooled predictors, with AUCs
of $0.66$–$0.65$ and AP $\approx 0.033$.

Urban ZIPs ($n \approx 8{,}293$) exhibit the highest raw AUCs, with internet and
computer access both exceeding $0.77$, although AP remains modest
($\approx 0.04$), consistent with lower positive prevalence \cite{strover2017,graves2021}. 
Rural ZIPs ($n \approx 13{,}883$) show lower AUCs overall ($0.65$–$0.71$) but higher AP
($\approx 0.068$), reflecting greater fragility prevalence and thus stronger
precision in the top of the ranked list \cite{harnack2019,saia2023}. 
Mixed ZIPs ($n \approx 3{,}065$) demonstrate the strongest \emph{relative} signal for education, with
AUC $=0.8651$ and AP $=0.228$; however, the smaller sample size warrants caution
when generalizing these gains.

Descriptive prevalence patterns are even starker. While Urban areas contribute
the bulk of absolute cases, Rural ZIPs dominate the lowest-uptake deciles:
$\approx 81\%$ of bottom-10\% ZIPs and $\approx 75\%$ of bottom-30\% ZIPs are
Rural. This asymmetry persists \emph{after} area-specific recalibration,
indicating that rural concentration of SNAP under-enrollment is structural rather
than an artifact of pooled quantiles \cite{harnack2019}.

\subsection{Hidden Fragility and Residual Diagnostics}
Residual-based diagnostics extended these findings. 
OLS residuals identified ZIPs with uptake far below linear expectations, 
some of which were not classified as fragile under the quantile thresholds. 
This suggests two forms of fragility: systemic—captured by high-poverty, 
low-uptake rules—and idiosyncratic, reflected in residual anomalies \cite{almond2011,hoynes2016}. 
The coexistence of these mechanisms points to both structural and 
context-specific barriers to SNAP participation.

Approximately 17{,}000 anomalous cases with SNAP counts exceeding poverty 
counts were also identified. These were retained but flagged to avoid 
inflating performance through selective deletion \cite{suttles2024}.

Detailed residual plots and anomaly tables supporting this analysis 
are provided in Appendix~\ref{app:residual}.

\subsection{Model Comparison: Logistic vs.~Tree-Based}
Table~\ref{tab:model-comparison} compares the leading logistic model with 
tree-based benchmarks. Logistic regression consistently outperformed 
alternative classifiers in both AUC and AP, while also producing 
interpretable coefficients. Tree-based models offered little incremental 
predictive power, consistent with the limited number of features \cite{suttles2024}.

\begin{table}[ht]
\centering
\caption{Comparison of best logistic regression, random forest, and gradient boosting models on P2. All values are percentages.}
\label{tab:model-comparison}
\resizebox{\textwidth}{!}{%
\begin{tabular}{lccccccc}
\hline
Model & AUC & AP & Precision & Recall & $F_1$ & Accuracy & Pre@1\% / Pre@5\% \\
\hline
Logistic (vehicle + HS diploma) & 71.1 & 9.8 & 14.1 & 20.2 & 16.6 & 92.3 & 14.8 / 13.6 \\
Random forest (all four)        & 66.3 & 6.4 & 7.3  & 2.9  & 4.1  & 94.9 &  --  /  --  \\
Gradient boosting (all four)    & 68.0 & 7.5 & 9.0  & 4.0  & 6.0  & 94.5 & -- / -- \\
\hline
\end{tabular}%
}
\end{table}

\begin{figure}[ht]
    \centering
    \includegraphics[width=0.7\linewidth]{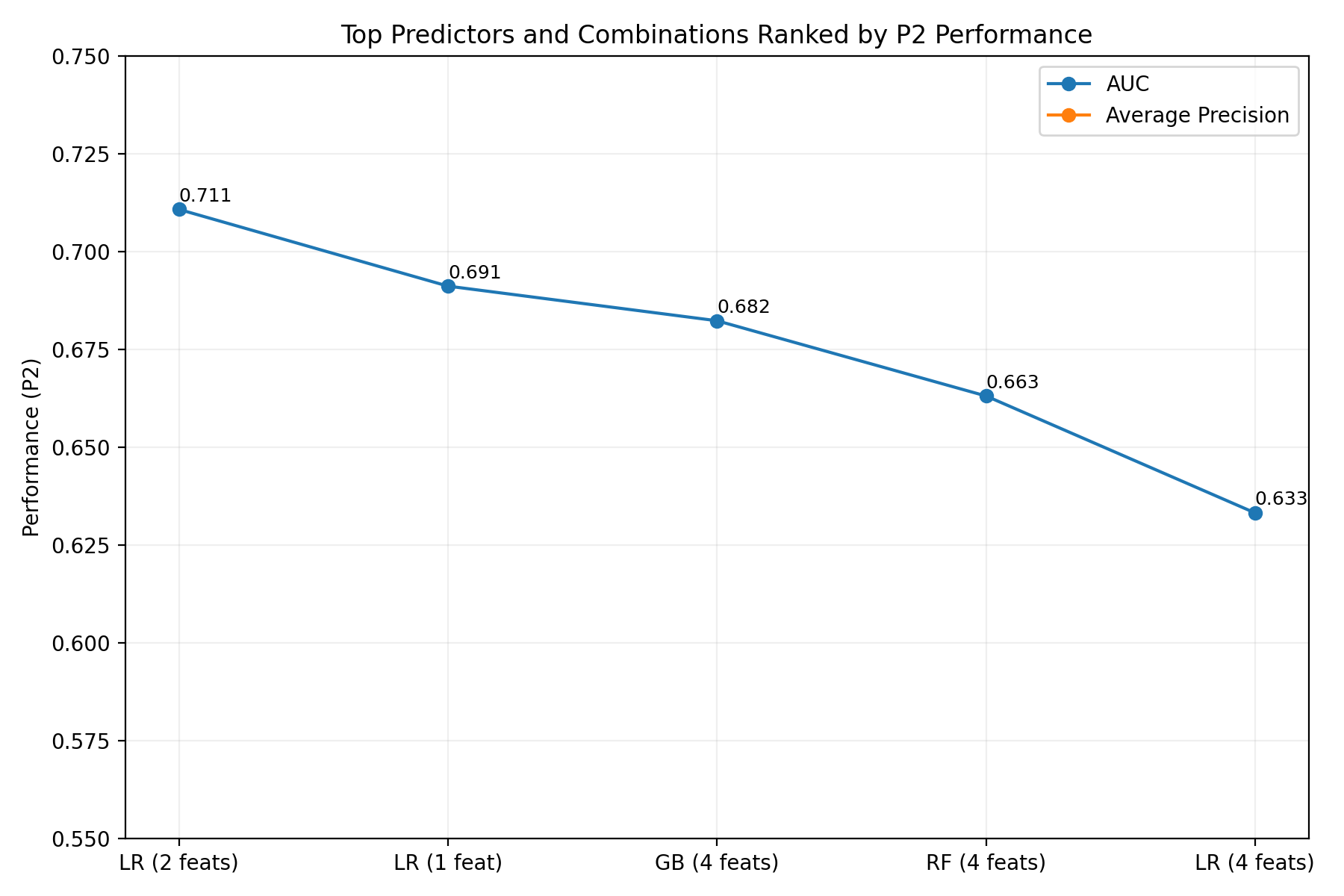}
    \caption{Comparison of top models on P2 ranked by AUC.
The two-feature logistic regression model (vehicle access + education) outperforms both gradient boosting and random forest models. This highlights the strength of simple, interpretable models in this context, especially given the limited feature set and class imbalance.}
    \label{fig:model_comparison_plot}
\end{figure}

\section{Conclusion}

This study investigated the ``SNAP Gap''—communities where poverty is high but 
Supplemental Nutrition Assistance Program (SNAP) participation remains 
unexpectedly low. These gaps are not statistical curiosities: they represent 
families who qualify for nutritional assistance but remain disconnected from it, 
leaving critical federal support unclaimed and food insecurity unaddressed. At a 
time when billions of dollars in SNAP benefits go unused each year \cite{cbpp2022}, 
identifying where and why under-enrollment occurs is a pressing policy challenge.

Across all specifications, logistic regression with vehicle access and education 
provided the clearest signal of fragility, with AUC $\approx 71$ and AP $\approx 10$. 
Tree-based methods, while more flexible, did not surpass logistic regression given 
the sparse predictor set. Fragility remains rare ($\pi_{\text{P1}} \approx 0.031$, 
$\pi_{\text{P2}} \approx 0.038$), underscoring the need for prevalence-adjusted 
metrics \cite{suttles2024}. Digital access and education emerged as consistent correlates of 
under-enrollment \cite{strover2017,graves2021,hoynes2016}, and residual diagnostics revealed an 
additional class of hidden fragility cases, suggesting avenues for qualitative follow-up.

By combining a transparent quantile-based definition with nationally available 
ZIP-level data, this study demonstrates that SNAP fragility is both measurable 
and structurally patterned. Rural communities are disproportionately affected, 
with vehicle access and educational attainment emerging as consistent predictors 
of under-enrollment \cite{harnack2019,saia2023}. These findings extend prior work on persistent 
poverty and the rural safety net by showing that barriers to program participation 
are not uniform, but cluster in ways that can be systematically flagged \cite{tiehen2013}. 
The modeling results also show that interpretable methods—simple logistic regression with a 
small set of socioeconomic indicators—can outperform more complex classifiers 
while remaining transparent to policymakers.

The broader implication is that gaps in safety net participation can be detected 
prospectively, not just described retrospectively. Training on one period and 
evaluating strictly on the next simulates how agencies might operationalize these 
insights for outreach, offering a practical framework for targeting scarce 
resources \cite{suttles2024}. The diagnostic use of residuals further highlights that fragility has 
both systemic and idiosyncratic forms, pointing to the need for combined 
statistical and qualitative investigation.

Ultimately, the ``SNAP Gap'' is not just a technical definition but a lens for 
understanding structural disconnection from public assistance. The framework 
developed here provides a scalable template for identifying under-enrollment, 
one that could be extended beyond SNAP to other programs where take-up lags 
eligibility. Closing these gaps means not only improving efficiency in benefit 
delivery, but also ensuring that vulnerable families receive the support that 
federal policy has already promised them \cite{berkowitz2017,ajph2019,jama2023}.

\section{Discussion, Policy Implications, and Future Research}

This study demonstrates that persistent under-enrollment in the Supplemental Nutrition Assistance Program (SNAP) reflects structural barriers within communities rather than individual behavior. Across more than 30,000 ZIP codes, two key factors—limited vehicle access and lower educational attainment—consistently predicted low participation in high-poverty areas. These findings suggest that the ``SNAP Gap'' is primarily driven by constraints on mobility and administrative literacy rather than by a lack of awareness or eligibility \cite{saia2023, hoynes2016}.

Transportation barriers appear to be the most consistent determinant of SNAP under-enrollment, particularly in rural areas where administrative offices and authorized retailers are often distant \cite{harnack2019}. The time, cost, and logistical challenges of traveling to apply for or redeem benefits substantially increase the real burden of participation. Educational attainment, meanwhile, functions as a proxy for procedural fluency—the ability to understand program rules, complete forms, and meet verification requirements \cite{suttles2024}. These patterns align with administrative burden theory \cite{moynihan2015}, which describes how learning, compliance, and psychological costs influence access to public benefits.

The patterns observed here illustrate how structural context shapes the administrative burden landscape. Areas with limited transportation and lower educational attainment face systematically higher learning and compliance costs, which translate into lower program take-up even when eligibility is constant. In this way, the SNAP Gap can be understood as a spatial manifestation of administrative burden, where infrastructure deficits—both physical and cognitive—concentrate the procedural challenges that the theory describes. Recognizing this connection helps clarify why simplifying forms or expanding digital access alone cannot close the gap without addressing underlying structural inequalities.

Earlier research emphasized digital access as a dominant limitation. However, the results of this study indicate that its importance has diminished as broadband and device ownership have become more widespread \cite{graves2021, pew2021}. The persistence of low SNAP participation despite improved digital connectivity underscores that access to transportation and education remains a more robust predictor of inclusion.

The results provide several policy insights. Agencies can use the SNAP Gap index to identify ZIP codes where poverty is high but program participation remains low, allowing for geographically targeted outreach and communication campaigns. Mobility-focused interventions such as mobile enrollment vans, on-site application stations in schools or libraries, and transportation vouchers could mitigate travel-related barriers \cite{saia2023}. Simplifying the administrative process through pre-filled renewal forms, plain-language materials, and digital auto-renewal tools may further reduce procedural burdens \cite{suttles2024}. In addition, the analytic framework presented here can be extended to other programs with similar under-enrollment challenges, such as WIC, Medicaid, or LIHEAP, to support a more unified approach to benefit access. Re-estimating the model annually with updated data would enable agencies to monitor changes in fragility and assess the impact of targeted outreach efforts over time.

Future research should integrate measures of race, ethnicity, and language isolation to examine whether structural fragility overlaps with demographic exclusion. Spatial dependence models could also clarify whether SNAP gaps cluster geographically or reflect broader regional disparities. Mixed-method approaches, including qualitative interviews and administrative case reviews, would deepen understanding of why some high-poverty ZIP codes continue to show low participation despite improvements in infrastructure. Future iterations of the model should also incorporate a broader set of contextual variables, such as housing cost burden, local labor market volatility, public transit coverage, and administrative office density, to capture additional dimensions of community fragility \cite{mckernan2017, ganong2019, alaimo2008}. Finally, linking this framework to outcomes such as health, education, or local food security would help assess how closing the SNAP Gap contributes to broader community well-being \cite{braveman2014, hoynes2016}.

In conclusion, addressing the SNAP Gap requires more than expanding eligibility criteria. It demands attention to the physical, informational, and procedural barriers that prevent households from accessing assistance. By identifying where these barriers are most severe, this study provides both a diagnostic tool and a practical framework for building a more equitable and effective safety net.

\appendix
\label{sec:appendix-area}
\section{Data Implementation Notes}
\begin{itemize}
    \item Raw inputs: PolicyMap CSVs (\texttt{pov\_fam}, \texttt{snap\_fam}, Plus4 metrics).
    \item HUD/USPS ZIP–tract crosswalk file and PolicyMap tract type CSVs for area designation.
\end{itemize}
\subsection{Cleaning and Assumptions}
Data were cleaned to enforce consistent structure:
\begin{itemize}
    \item Sentinel codes recoded to NA; percentages clipped to $[0,100]$.
    \item Invalid or duplicate rows dropped or averaged.
    \item SNAP $>$ poverty anomalies flagged but retained.
\end{itemize}

\subsection{Sample Restrictions and Exclusions}
\begin{itemize}
    \item Valid years: 2014--2023.
    \item Eligibility: $p_z \geq 0.15$, $s_z$ finite, all Plus4 predictors non-missing.
    \item Zero uptake coerced to NA; excluded from target construction.
    \item ``Unknown'' ZIPs excluded from modeling.
\end{itemize}

\subsection{Data Integration}
\begin{itemize}
    \item HUD crosswalk used to aggregate tract-level PolicyMap data to ZCTAs.
    \item ZIP/ZCTA codes normalized to five digits for consistent joins.
    \item Area designations fixed from 2019 categories and applied to both periods.
\end{itemize}

\section{Methods Implementation Notes}
\begin{itemize}
    \item Cleaning scripts: \texttt{PM\_full\_fam\_cleaner.py}, \texttt{build\_zip\_area\_designation.py}.
    \item modeling scripts: \texttt{backtester.py} for logistic/forest/boosting models.
\end{itemize}
\subsection{Learning Setup}
Three model families are benchmarked:
\begin{enumerate}
    \item Logistic regression with standardized predictors and balanced class weights. 
    Hyperparameters explored include regularization strength $C$ across coarse grids.
    \item Random forests varying in tree depth and number of estimators, 
    trained with class weighting to account for rare positives.
    \item Gradient boosting machines with tuned learning rates, depths, and estimators.
\end{enumerate}
Hyperparameters are selected via stratified 5-fold cross-validation within P1, 
ensuring that model choice does not overfit the small positive class. 
To improve probability calibration, isotonic regression is applied to 
cross-validated predictions, yielding calibrated scores $\tilde{p}_z$.

\subsection{Decision Rule}
Final binary predictions adopt a prevalence-anchored rule:
\[
\hat{y}_z = \mathbf{1}\{ \tilde{p}_z \ge \pi_{\text{P1}} \},
\]
where $\pi_{\text{P1}}$ is the positive prevalence in training. 
This choice ties the threshold to the rarity of fragility, rather than 
arbitrary values like 0.5, and ensures comparability across years and models.

\subsection{Evaluation Metrics}
Performance is assessed by a suite of metrics. ROC AUC measures general 
discrimination, but given the $\sim$3--4\% prevalence, precision--recall 
curves and Average Precision (AP) are more informative. Precision, recall, and 
$F_1$ are reported at the prevalence-anchored decision rule. Accuracy is reported 
but interpreted cautiously due to class imbalance. Precision@1\% and Precision@5\% 
capture performance in the highest-risk ZIPs. Calibration quality is evaluated 
graphically via reliability curves. All metrics are reported on 5,466 eligible 
ZIPs in P2 ($\pi_{\text{P2}} \approx 3.8\%$).

\subsection{Geographic and Temporal Partitioning}
Models are re-estimated within urban, rural, and mixed subsets, each with 
their own cutpoints $(\tau_{\text{hi}},\tau_{\text{lo}})$. 
This accounts for the possibility that fragility manifests differently 
depending on settlement type. Temporal diagnostics apply the same logic 
year by year from 2014--2023. For forests, feature importance is quantified 
by permutation:
\[
\operatorname{Imp}(j) = 
\mathbb{E}\!\left[ M(f_\theta(X)) - M(f_\theta(X_{\pi_j})) \right],
\]
where $M(\cdot)$ is ROC AUC or AP, $X$ is the design matrix, and $X_{\pi_j}$ 
is the same matrix with predictor $j$ permuted.

\subsection{Fixed Design Choices}
Quantile levels were fixed at HI\_Q=0.70 and LO\_Q=0.10; the poverty floor was 
set at $\underline{p}=0.15$. Area cutoffs classify ZIPs as urban if at least 
80\% urban, rural if at most 20\%, and mixed otherwise. 
Alternative quantiles or thresholds were not explored in this study, 
focusing instead on consistent operational definitions. 
Records with SNAP $>$ poverty were flagged but retained to avoid artificially 
inflating performance by discarding anomalies.

\section{Area-Level Analysis}
\label{app:area}

This appendix provides additional detail on area designations, sample distributions, and model performance by settlement type. It supplements Section~3.7 (“Area Context and Stratified Modeling”) in the main text.

\subsection{Area Types and Thresholds}
Area designations were derived by joining PolicyMap tract-level classifications with the HUD ZIP--tract crosswalk. 
Residential weights (\texttt{RES\_RATIO}) were summed by tract status and thresholded at 0.80/0.20: 
ZIPs with $\geq 80\%$ urban share were labeled Urban, $\leq 20\%$ urban share as Rural, 
and intermediate values as Mixed; cases with missing or indeterminate status were labeled Unknown. 
These designations were merged onto every P1 and P2 record prior to modeling.

Quantile thresholds were recomputed for each area to ensure local comparability. 
In P2, the 70th-percentile poverty thresholds were 24.4\% (Urban), 25.2\% (Rural), 21.1\% (Mixed), 
and 38.2\% (Unknown).  Bottom-decile uptake thresholds varied substantially: 
0.881 (Urban), 0.333 (Rural), 0.742 (Mixed), and 0.495 (Unknown). 

\subsection{Area-Specific Modeling}
For each area subset (Urban, Rural, Mixed), model fitting proceeded separately. 
Univariate models used a stratified 70/30 train--test split with Youden-threshold classification, 
while multivariate specifications (logistic regression and tuned random forest) employed 3-fold 
stratified cross-validation. Logistic thresholds were aligned to area-specific prevalence levels. 
For example, in P2 the optimal decision cutpoints were 0.1823 for the All sample, 0.0955 for Urban, 
0.2500 for Rural, and 0.1385 for Mixed ZIPs. 
Permutation importances from random forests were computed within each subset to capture 
area-specific predictor relevance.

\subsection{Sample Totals and Prevalence}
\begin{table}[H]
\centering
\caption{Sample Totals and Prevalence (P2 and P1)}
\begin{tabular}{lcccc}
\toprule
\textbf{Area Type} & \textbf{Rows (P2)} & \textbf{Eligible (P2)} & \textbf{Eligible (P1)} & \textbf{Pos.~Rate (P2)} \\
\midrule
Urban   & 8,966  & 1,505 & 8,874  & 3.1\% \\
Rural   & 17,840 & 2,875 & 17,613 & 3.6\% \\
Mixed   & 3,148  & 428   & 3,140  & 4.7\% \\
Unknown & 3,837  & 658   & 3,516  & 1.4\% \\
\midrule
\textbf{Overall} & 33,791 & 5,466 & 33,143 & P1: 3.1\%, P2: 3.8\% \\
\bottomrule
\end{tabular}
\end{table}

\subsection{rea-Specific Cutoffs}
\begin{table}[H]
\centering
\caption{P2 Area-Specific Poverty and Uptake Cutoffs}
\begin{tabular}{lcc}
\toprule
\textbf{Area Type} & \textbf{70th-Percentile Poverty Threshold} & \textbf{Bottom-10\% Uptake Cutoff} \\
\midrule
Urban   & 24.4\% & 0.881 \\
Rural   & 25.2\% & 0.333 \\
Mixed   & 21.1\% & 0.742 \\
Unknown & 38.2\% & 0.495 \\
\bottomrule
\end{tabular}
\end{table}

\subsection{Distribution of Fragile ZIPs}
\begin{table}[H]
\centering
\caption{Distribution of Fragile ZIPs (P2)}
\begin{tabular}{lcccc}
\toprule
\textbf{Group} & \textbf{Rural} & \textbf{Unknown} & \textbf{Urban} & \textbf{Mixed} \\
\midrule
Bottom 10\% & 169 (81.3\%) & 28 (13.5\%) & 9 (4.3\%) & 2 (1.0\%) \\
Bottom 30\% & 380 (74.7\%) & 75 (14.7\%) & 45 (8.8\%) & 9 (1.8\%) \\
\bottomrule
\end{tabular}
\end{table}

\subsection{Area-Specific Performance}
\begin{table}[H]
\centering
\caption{Best Predictors by Area Type (P2)}
\begin{tabular}{lccc}
\toprule
\textbf{Area Type} & \textbf{Best Predictor} & \textbf{AUC} & \textbf{AP} \\
\midrule
All ZIPs  & \% no internet     & 0.663 & 0.033 \\
Urban     & \% no internet     & 0.782 & 0.040 \\
Rural     & \% no internet     & 0.711 & 0.068 \\
Mixed     & \% HS diploma only & 0.865 & 0.228 \\
\bottomrule
\end{tabular}
\end{table}

\section{Model Diagnostics}
\label{app:diagnostics}

\subsection{Logistic Coefficients}
To interpret model behavior beyond overall performance, we examined the 
coefficients of the best-performing logistic regression specification. 
Standardized coefficients provide a sense of which predictors most strongly 
contributed to identifying low-uptake ZIPs, and in which direction. Negative 
coefficients indicate predictors associated with higher fragility (lower SNAP 
uptake given poverty), while positive coefficients suggest protective effects.

\subsection{Calibration Diagnostics}
A calibration curve was generated for the univariate logistic model using 
\% no vehicle as the predictor, after applying isotonic regression to 
cross-validated probabilities. This plot illustrates that predicted 
fragility probabilities align closely with observed rates, confirming 
excellent calibration for the primary explanatory feature..

\begin{figure}[H]
    \centering
    \includegraphics[width=0.7\linewidth]{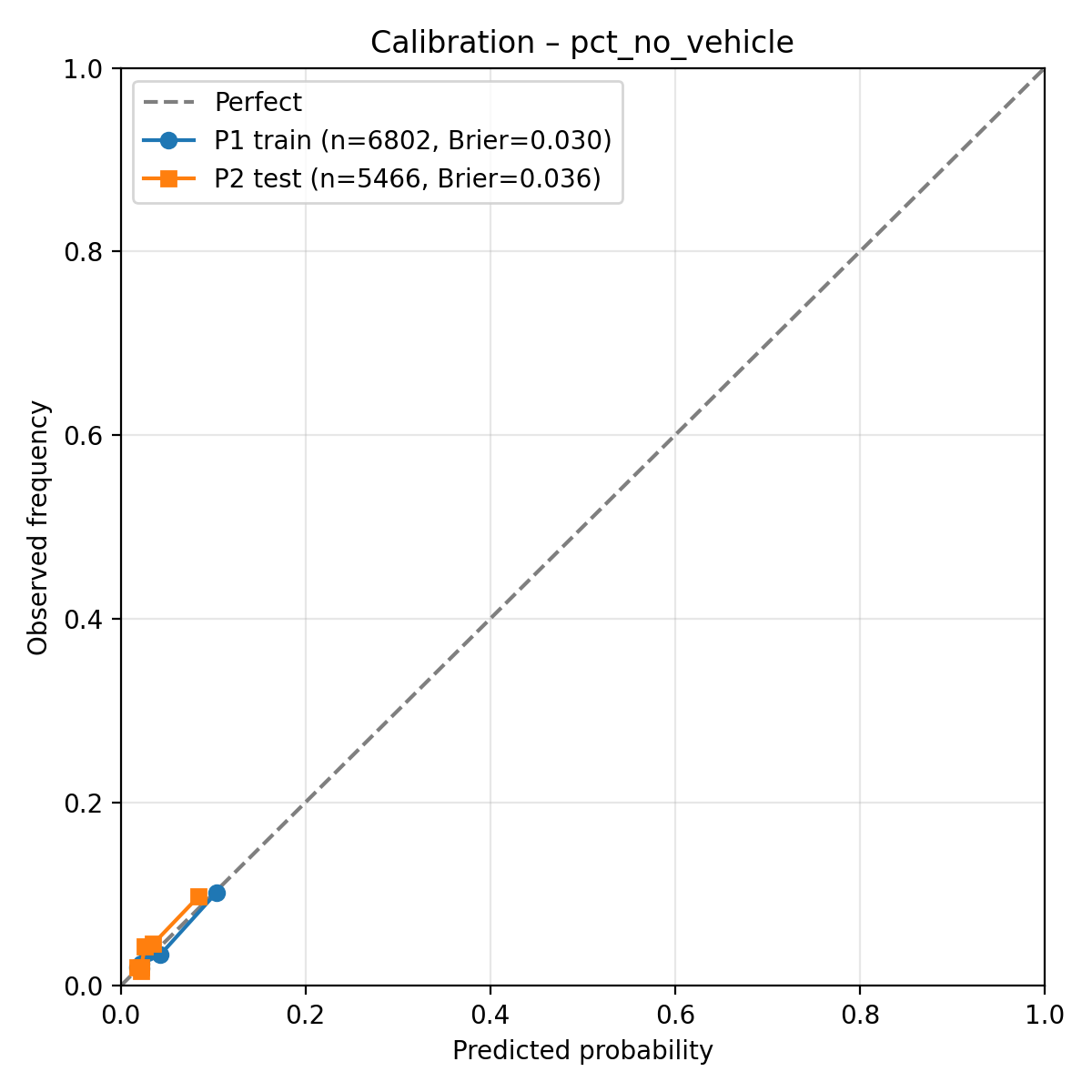}
    \caption{Calibration plot for \% no vehicle (univariate logistic model).
    The isotonic-corrected probabilities align closely with the ideal 45-degree line, 
    especially in the upper decile. This indicates excellent calibration—predicted 
    probabilities reflect actual risk well—reinforcing vehicle access as a robust predictor.}
    \label{fig:calibration_pct_no_vehicle}
\end{figure}

\section{Residual and Quality Control Checks}
\label{app:residual}

\subsection{OLS Residual Anomalies}
OLS regression residuals flagged ZIPs with anomalously low SNAP uptake 
relative to poverty counts. These residual-flagged cases were not always 
captured by quantile cutoffs, indicating that fragility can manifest in 
distinct ways beyond the high-poverty, low-uptake rule.

\subsection{Cases in which SNAP Exceeds Poverty}
A set of approximately 17{,}000 anomalous cases exhibited SNAP counts 
exceeding estimated poverty counts. These may reflect reporting 
inconsistencies, definitional differences between poverty and SNAP 
household measures, or policy factors such as categorical eligibility 
extensions to households slightly above the poverty line.

\bibliographystyle{plainnat}  
\bibliography{snapgapfinal}   
\end{document}